
\input harvmac

\Title{hep-th/9510138 YCTP-P14-95}{\vbox{\centerline{Discontinuous BPS spectra
in $N = 2$ gauge theory}}}

\centerline{M{\aa}ns Henningson}
\bigskip{\it \centerline{Yale University}\centerline{Department of
Physics}\centerline{P. O. Box 208120}\centerline{New Haven, CT 06520-8120,
USA}}\centerline{mans@genesis5.physics.yale.edu}

\vskip 40mm \centerline{\bf Abstract}
We consider the spectrum of BPS saturated states in $N = 2$ gauge theories in
four dimensions. This spectrum may be discontinuous across real codimension one
submanifolds of marginal stability in the moduli space of vacua. An example,
which can be treated with semiclassical methods in the weak coupling limit, is
the decay of quark-soliton bound states. For a quark and a soliton of
electric-magnetic charge vectors $Q$ and $Q^\prime$ respectively, we find that
as the manifold of marginal stability is crossed, the number of soliton states
changes by a factor of $2^{Q \cdot Q^\prime}$, where the dot denotes the
symplectic product.

\Date{October 1995}

\newsec{Introduction}
An important feature of $N = 2$ supersymmetric theories in four space-time
dimensions is the existence of a central charge $Z$ in the supersymmetry
algebra \ref\HLS{
R. Haag, J. T. Lopuszanski and M. Sohnius, {\it Nucl. Phys.} {\bf B88} (1975)
257.
}:
\eqn\algebra{
\{ Q_\alpha^A, Q_\beta^B \} = \epsilon_{\alpha \beta} \epsilon^{AB} Z.
}
The central charge is in general a linear combination of conserved abelian
charges with coefficients that depend on the vacuum moduli and the parameters
of the Lagrangian. In a unitary representation of supersymmetry, the mass $m$
of a state is bounded \ref\WO{
E. Witten and D. Olive, {\it Phys. Lett.} {\bf 78B} (1978) 97.
} by
\eqn\bound{
m \geq | Z |.
}
If we limit ourselves to maximum spin $1$ (i.e. we do not consider supergravity
theories) we have the following representations of the supersymmetry algebra
\algebra: When the inequality \bound\ is not saturated, an irreducible
representation contains five scalar, four spinor and one vector particle, i.e.
a total of sixteen helicity states. For a so called BPS representation, which
saturates \bound, there are two possibilities; either two scalar and one spinor
particle, or one scalar, two spinor and one vector particle, for a total of
four or eight helicity states respectively.

Particles in a BPS representation enjoy the following stability property: For a
BPS state with central charge $Z$ and mass $m$ to decay into states with
central charges $Z_1$ and $Z_2$ and masses $m_1$ and $m_2$, we must have $Z =
Z_1 + Z_2$ by charge conservation. But this means that $m = | Z | \leq | Z_1 |
+ | Z_2 | \leq m_1 + m_2$ by the triangle inequality, so this decay cannot take
place. These inequalities are saturated exactly when the would-be decay
products are also BPS states and the central charges $Z_1$ and $Z_2$ are
aligned in the complex plane. This will happen on a real codimension one
subvariety of `marginal stability' in the moduli space of vacua. As explained
in \ref\SWI{
N. Seiberg and E. Witten, {\it Nucl. Phys.} {\bf B426} (1994) 19.
}, the spectrum of BPS one-particle states must sometimes be discontinuous
across such
a boundary. This phenomenon is well understood in two space-time dimensions
\ref\CVII{
S. Cecotti and C. Vafa, {\it Comm. Math. Phys.} {\bf 158} (1993) 569.
}, where the number of particles appearing or disappearing in this way can be
calculated from Picard-Lefschetz theory in the case of Landau-Ginzburg
theories, or from the ${\rm Tr} (-1)^F F \exp (-\beta H)$ index \ref\CFIV{
S. Cecotti, P. Fendley, K. Intriligator and C. Vafa, {\it Nucl. Phys.} {\bf
B386} (1992) 405.
} and the $t t^*$ topological-anti-topological fusion equations \ref\CVI{
S. Cecotti and C. Vafa, {\it Nucl. Phys.} {\bf B367} (1991) 359.
} for general $N = 2$ theories.

The purpose of this article is to study such discontinuities in the context of
$N = 2$ gauge theories in four dimensions. In the simplest example, the $N = 2$
$SU(2)$ gauge theory without extra matter \SWI, there are only two conserved
abelian charges, namely the electric and magnetic charge for the unbroken
$U(1)$ gauge group. Consequently, there is a single curve of marginal stability
where their coefficients in the central charge are aligned. Unfortunately, this
curve lies entirely in the strong coupling regime \ref\AFS{
P. C. Argyres, A. E. Faraggi and A. D. Shapere, hep-th/9505190.
}, where it is difficult to exhibit the discontinuity of the spectrum
explicitly. In more complicated examples, i.e. with larger gauge groups and/or
extra matter, there are more conserved abelian charges and the manifold of
marginal stability depends on the quantum numbers of the particles involved in
the decay. (The set of all such curves, regardless of which particles actually
exist in the spectrum, is in fact dense in the moduli space.) More important
for our purposes is that models with extra matter have dimensionful parameters,
namely the bare masses of the matter fields. When such a parameter is large
compared to the dynamically generated scale of the theory, marginal stability
might occur also at weak coupling, where semiclassical methods are applicable.

This paper is organized as follows: In section two, we quickly review some
aspects of $N = 2$ gauge theories, and in section three, we discuss their BPS
spectra at weak coupling. In section four, we calculate the number of
quark-soliton bound states that disappear as a curve of marginal stability is
crossed. In section five, we consider the case of an $SU(2)$ gauge group in
somewhat more detail.

\newsec{$N = 2$ supersymmetric gauge theories}
The field content of our model consists of an $N = 2$ gluon vector multiplet
and an $N = 2$ quark hypermultiplet. In terms of $N = 1$ superfields, the
vector multiplet may be decomposed as an $N = 1$ vector multiplet $V$ and an $N
= 1$ chiral multiplet $\Phi$, both transforming in the adjoint of the gauge
group $G$. The hypermultiplet consists of two $N = 1$ chiral multiplets $Q$ and
$\tilde{Q}$ transforming in complex conjugate (in general reducible)
representations $R$ and $\bar{R}$ under $G$. To have an asymptotically free or
scale invariant model, the index of the representation $R$ must be less than or
equal to the index of the adjoint representation of $G$. The action of the
model is completely determined by $N = 2$ supersymmetry to be of the form
\eqn\action{
{\cal L} = \int d^2 \theta d^2 \bar{\theta} \left( {\rm Im} \, \tau \,
\Phi^\dagger e^V \Phi + Q^\dagger e^V Q + \tilde{Q}^\dagger e^V \tilde{Q}
\right) + \int d^2 \theta \left ( \tau \, W^\alpha W_\alpha + \tilde{Q} (\Phi +
m) Q \right) + {\rm c.c.},
}
where $m$ gives the bare masses of the irreducible components of the
hypermultiplet, and the gauge coupling constant and the theta-parameter are
combined as $\tau = {\theta \over 2 \pi} + i {4 \pi \over g^2}$. (We use the
conventions of  \ref\WB{
J. Wess and J. Bagger, {\it Supersymmetry and Supergravity}, (Princeton
University Press 1983).
} for $N = 1$ superspace. Note that our normalization of $m$ differs from that
of \ref\SWII{
N. Seiberg and E. Witten, {\it Nucl. Phys.} {\bf B431} (1994) 484.
}\ by a factor of $\sqrt{2}$.) The classical action \action\ with vanishing
bare masses is invariant under an $U(1)_{\cal R}$ symmetry such that the ${\cal
R}$-characters of $\Phi$, $W_\alpha$, $Q$ and $\tilde{Q}$ are $1$, $1/2$, $0$
and $0$ respectively. The masses $m$ and the dynamically generated scale
$\Lambda$ in an asymptotically free theory break this symmetry, but we may
restore it by assigning ${\cal R}$-character $1$ to $m$ and $\Lambda$.

The theory has a moduli space of inequivalent vacuum states. We will consider
the Coulomb branch of this moduli space in which only the lowest component
$\phi$ of the chiral superfield $\Phi$ has a non-vanishing vacuum expectation
value, subject to the constraint $[  \phi  ,  \phi^\dagger  ] = 0$. The theory
will be weakly coupled when $< \! \phi \! >$ is large compared to the scale
$\Lambda$. At a generic point on this branch, the gauge symmetry is
spontaneously broken to its maximal abelian subgroup $U(1)^r$, where $r = {\rm
rank} (G)$, and we have a vector $q$ of corresponding electric charges. As
usual in spontaneously broken gauge theories where the exact symmetry group
contains abelian factors, there will also be a vector of magnetic charges $g$
in the theory. In general, these charges are not well-defined over the moduli
space of vacua; if we assemble $q$ and $g$ to an electric-magnetic charge
vector $Q = (q, g)$, this ambiguity amounts to the action of an element of
$Sp(2 r, {\bf Z})$ on $Q$ as we encircle a singularity \SWI. Finally, there is
a set of conserved quark number charges $S$, one for each irreducible component
of the representation $R$. These charges can pick up contributions proportional
to $q$ and $g$ as we encircle a singularity of the moduli space \SWII.

\newsec{The BPS spectrum}
The central charge $Z$ for a state is a linear combination of the electric,
magnetic and quark number charges:
\eqn\Z{
Z = a \cdot q + a_D \cdot g + m \cdot S,
}
where classically  $a$ is given by the eigenvalues of $\phi$ and $a_D = \tau
a$, but these coefficients get modified at the quantum level. The formula \Z\
is derived by explicitly constructing the supercharges and calculating their
Poisson bracket \WO\ \SWI\ \SWII. (The contributions from electric and quark
number charges are most easily checked by calculating the anticommutator of two
supersymmetry transformations on the component fields of the vector and
hypermultiplets.)

We will now determine the spectrum of BPS states in the theory, at least at
weak coupling (i.e. for $< \! \phi \! >$ large compared to $\Lambda$ in
asymptotically free theories or ${\rm Im} \, \tau$ large) where semiclassical
methods can be trusted. We start with the elementary field quanta. The fields
of the $N = 2$ vector multiplet fill out a BPS representation of maximum spin
$1$. Indeed, the bound \bound\ is saturated by the usual Higgs formula for the
mass of gauge bosons in a situation with spontaneous symmetry breaking \ref\MO{
C. Montonen and D. Olive, {\it Phys. Lett.} {\bf 72B} (1977) 117.
}. The $N = 2$ hypermultiplet must also give rise to BPS representations, since
it only contains fields of maximum spin $1/2$. The central charges of the
different particles are given by the eigenvalues of the matrix $\phi + m$, and
the masses, which can be read off from the action \action, again saturate
\bound.

The theory also contains magnetically charged solitonic excitations with
non-vanishing expectation values of $\phi$ and the gauge field $A_\mu$. The
bound \bound\ is saturated by the Bogomolny-Prasad-Sommerfield limit \ref\B{
E. B. Bogomolny, {\it Sov. J. Nucl. Phys.} {\bf 24} (1976) 449.
} \ref\PS{
M. K. Prasad and C. M. Sommerfield, {\it Phys. Rev. Lett.} {\bf 35} (1975) 760.
} of monopoles and dyons \ref\tH{
G. 't Hooft, {\it Nucl. Phys.} {\bf B79} (1974) 276.
}\nref\Po{
A. Polyakov, {\it JETP Lett.} {\bf 20} (1974) 194.
} -- \ref\JZ{
B. Julia and A. Zee, {\it Phys. Rev.} {\bf D11} (1975) 2227.
}. Such a configuration breaks half of the supersymmetries \ref\Os{
H. Osborn, {\it Phys. Lett.} {\bf 83B} (1979) 321.
}. Acting with the broken supersymmetry generators on a state of spin $0$, we
can create two helicity states transforming as spin $1/2$ and one more spin $0$
state, i.e. a BPS multiplet of maximum spin $1/2$. We should also include the
CPT conjugate states of opposite electric and magnetic charges for a total of
eight helicity states. After an electric-magnetic duality transformation, these
states could be described by a hypermultiplet coupled to the dual photons in a
$U(1)^r$ gauge theory \SWI.

Another type of BPS states can be thought of as bound states in a quark-soliton
system. The fermionic fields in the $(Q ,\tilde{Q})$ hypermultiplet make up a
Dirac spinor $\psi$ and a conjugate spinor $\bar{\psi} = \psi^\dagger \gamma^0$
transforming in the $R$ and $\bar{R}$ representations respectively. The
equation of motion for these fields derived from \action\ in a fixed $A_\mu$
and $\phi$ background reads
\eqn\Diracequation{
\left( i \gamma^\mu D_\mu  - {\rm Re} (\phi + m) + \gamma^5 {\rm Im} (\phi + m)
 \right) \psi = 0,
}
where we have decomposed the matrix $\phi +m$ in its Hermitian and
anti-Hermitian parts. A convenient basis for the Dirac matrices is
\eqn\Diracmatrices{
\gamma^0 = i \left( \matrix{ 0 & 1 \cr -1 & 0} \right) \;\;\;\;\;\;\;\;\;\;\;\;
\gamma^i = i \left( \matrix{ \sigma^i & 0 \cr 0 & -\sigma^i} \right) \;\;\;\; i
= 1, 2, 3 \;\;\;\;\;\;\;\;\;\;\;\; \gamma^5 = i \left( \matrix{ 0 & -1 \cr -1 &
0} \right),
}
where the $\sigma^i$ are the Pauli matrices. We decompose the Dirac spinor
accordingly as $\psi = \left( \matrix{ \psi^+ \cr \psi^-} \right)$. Choosing
the gauge $A_0 = 0$, replacing $-i \partial_0$ by the energy eigenvalue $E$ and
multiplying equation \Diracequation\ by $\gamma^0$ from the left, we get
\eqn\Schrodinger{
\eqalign{
\left( \matrix{ {\rm Im} (\phi + m) - E & L \cr L^\dagger & - {\rm Im} (\phi +
m) - E } \right) \left( \matrix{ \psi^+ \cr \psi^- } \right) = 0,
}
}
where the operator $L$ is given by
\eqn\Loperator{
L = i \sigma^i D_i - i {\rm Re} (\phi + m).
}
Each normalizable solution to \Schrodinger\ doubles the number of BPS solitons,
since a fermionic state can be either empty or occupied.

\newsec{The discontinuity}
It might seem difficult to find solutions to \Schrodinger. However, we know
that the spectrum of BPS states cannot change unless we cross a curve of
marginal stability. To determine the number of bound states that appear or
disappear across such a curve, we choose a
specific point on the curve where $\phi$ is Hermitian up to a phase, which we
can rotate away by a $U(1)_{\cal R}$ transformation so that ${\rm Im} \, \phi =
0$. If we now have a solution to $L \psi^- = 0$, a solution of \Schrodinger\
would be to take $\psi^+ = 0$ and $E = -{\rm Im} \, m$. Similarly, a solution
to $L^\dagger \psi^+ = 0$ gives a solution to \Schrodinger\ with $\psi^- = 0$
and $E = {\rm Im} \, m$. Note that since the central charges of the quark
components are given by the eigenvalues of $\phi + m$ and the central charge
$Z$ of a magnetically charged soliton configuration in the weak coupling limit
is a large imaginary number, the mass difference between a quark-soliton bound
state and the soliton state should indeed equal $| {\rm Im} \, m|$.

We have thus found a space of solutions to \Schrodinger\ of dimension ${\rm dim
\, Ker} (L) + {\rm dim \, Ker} (L^\dagger)$. In general, this quantity is
difficult to determine, but a lower bound is given by the absolute value of
${\rm index} (L) = {\rm dim \, Ker} (L) - {\rm dim \, Ker} (L^\dagger)$. Now we
can use the three-dimensional version of the Callias index theorem \ref\Ca{
C. Callias, {\it Comm. Math. Phys.} {\bf 62} (1978) 213.
}:
\eqn\Callias{
{\rm index} (L) = -{1 \over 16 \pi i} \int {\rm Tr} \left( U dU \wedge dU
\right),
}
where $U = -| {\rm Re} (\phi + m) |^{-1} {\rm Re} (\phi + m)$ and the integral
is taken over a two-sphere at spatial infinity. The matrix $U$ is only
well-defined if the eigenvalues of ${\rm Re} (\phi + m)$ are non-zero; this is
in fact a necessary condition for $L$ to be Fredholm so that ${\rm index} (L)$
is well-defined. As long as we stay away from zero eigenvalues of ${\rm Re}
(\phi + m)$, ${\rm index} (L)$ depends continuously on $\phi + m$ and is
therefore a constant.

However, the index can change when an eigenvalue changes sign. Indeed, the
central charges of the quark components are given by the eigenvalues of $\phi +
m$. With ${\rm Im} \, \phi = 0$, there will be a purely imaginary central
charge whenever ${\rm Re} (\phi + m)$ has a zero eigenvalue. This aligns with
the central charge of a magnetically charged soliton in the weak coupling
limit, and a discontinuity in the BPS spectrum is therefore possible. For ${\rm
Re} \, m$ large compared to the scale $\Lambda$, the discontinuity will take
place in the weak coupling regime, thus justifying our reliance on
semiclassical methods. We  can decompose ${\rm Re} (\phi + m)$ as
\eqn\project{
{\rm Re} (\phi + m) = \sum_A \lambda_A \, \chi^{}_A \chi^\dagger_A,
}
where the $\lambda_A$ are the real eigenvalues of ${\rm Re} (\phi + m)$ and the
$\chi_A$ the corresponding eigenvectors normalized up to an arbitrary phase by
$\chi^\dagger_A \chi^{}_B = \delta_{A B}$ and fulfilling the completeness
relation $\sum_A \chi^{}_A \chi^\dagger_A = 1$. The matrix $U$ in \Callias\ is
then given by
\eqn\Umatrix{
U = \sum_A {\rm sign} (\lambda_A) \, \chi^{}_A \chi^\dagger_A.
}
We consider a situation where one of the eigenvalues, $\lambda_0$ say, changes
sign. A short calculation gives
\eqn\difference{
\lim_{\lambda_0 \rightarrow 0+} {\rm Tr} \left( U dU \wedge dU \right) -
\lim_{\lambda_0 \rightarrow 0-} {\rm Tr} \left( U dU \wedge dU \right) = 4 d
\chi_0^\dagger \wedge d \chi_0^{}.
}
An important point is now that $\chi_0$ is in general not a globally defined
function over $S^2$ but a section of the line-bundle $V$ of eigenvectors of
$\phi$ with eigenvalue $- {\rm Re} \, m$. The quantity on the right-hand side
of \difference\ is globally well-defined, though, and we can calculate the jump
in the index as
\eqn\jump{
\Delta {\rm index} (L) = - {1 \over 2 \pi i} \int d \chi_0^\dagger \wedge d
\chi_0^{} = c_1 (V) [ S^2 ].
}
This is of course an integer as it should be, since the first Chern class of
$V$ is an element of $H^2 (S^2, {\bf Z})$. In physical terms, this amounts to
the Dirac quantization condition: The wave function of a particle with
electric-magnetic charge vector $Q$ in the presence of  charges $Q^\prime$ at
the origin is a section of a line-bundle over ${\bf R}^3 - \{ 0 \}$, the first
Chern number of which equals the integer symplectic product $Q \cdot Q^\prime$.
The number of solitonic states thus changes by a factor of $2^{Q \cdot
Q^\prime}$ as a curve of marginal stability is crossed. This number is
invariant under the action of $Sp(2 r, {\bf Z})$ on $Q$ and $Q^\prime$ and thus
well-defined on the moduli space of vacua. One should note that although we
have considered fundamental quarks in a soliton background in the weak coupling
regime, the result has a wider validity. For example, as the mass parameters of
the model are turned off, the elementary quarks may be continuously changed to
magnetically charged solitons, and curves of marginal stability may move from
the weak coupling to the strong coupling regime \SWII. The discontinuities of
the BPS spectrum are unchanged, though. It would be interesting to check the
validity of this result in cases which can not be treated at weak coupling, and
also for dyon-dyon bound states.

\newsec{The $SU(2)$ case}
The case of an $SU(2)$ gauge group is of particular importance, since embedded
$SU(2)$ monopoles and dyons seem to be the fundamental one-particle solitons
also for larger gauge groups \ref\We{
E. Weinberg, {\it Nucl. Phys.} {\bf B167} (1979) 500.
}.

A gauge invariant parametrization of the moduli space of vacua is given by the
vacuum expectation value of $u = {1 \over 2} {\rm Tr} (\phi^2)$. By using the
$U(1)_{\cal R}$ symmetry, we can take $u$ to be real and positive. The monopole
configuration \tH\ \Po\ is of the form
\eqn\monopole{
\phi = \hat{x} \cdot \sigma \, \phi (r) \;\;\;\;\;\;\;\;\;\;\;\; A_0 = 0
\;\;\;\;\;\; A_i = \epsilon_{i j k} \sigma^j \hat{x}^k A(r),
}
where $r = (x \cdot x)^{1/2}$, $\hat{x}^i = r^{-1} x^i$ and $\phi(r)$ is real
and positive. We have $\phi(r)  \rightarrow \sqrt{u}$ and $A(r) \rightarrow 0$
as $r \rightarrow \infty$. This background is invariant under rotations
generated by $J_i = L_i + S_i + T_i$, where $L_i$, $S_i$ and $T_i$ are the
generators of orbital angular momentum, spin and isospin respectively.

With an $SU(2)$ gauge group, asymptotic freedom or scale invariance permits us
to consider quarks in (up to four copies of) the doublet representation or (a
single copy of) the triplet representation. The Callias formula \Callias\ then
predicts one or two normalizable modes respectively, which have been explicitly
constructed in the case of zero bare quark mass \ref\JR{
R. Jackiw and C. Rebbi, {\it Phys. Rev.} {\bf D13} (1976) 3398.
}. The two modes for a triplet quark transform as spin $1/2$ under $J_i$,
leading to solitons in BPS representations of maximum spin $1$. Doublets and
triplets of $SU(2)$ are the only representations that arise also when we embed
a monopole in an asymptotically free or scale invariant theory with a larger
gauge group. Triplets arise when the quarks are in the adjoint representation
of the gauge group (i.e. in $N = 4$ theories), but also in $SU(n)$ theories
with quarks in the symmetric product of two fundamental representations.

Here we will consider the case of isodoublet quarks with non-zero bare mass in
somewhat more detail. The lowest mode should share the symmetry of the monopole
background under rotations generated by $J_i$. This leads to the Ansatz
\eqn\Ansatz{
(\psi^-)_\alpha{}^\kappa = \delta_\alpha{}^\kappa \psi^-_s (r) + (\hat{x} \cdot
\sigma)_\alpha{}^\kappa \psi^-_v (r),
}
where $\alpha = 1, 2$ and $\kappa = 1, 2$ are spinor and $SU(2)$ doublet
indices respectively. With $L$ as in \Loperator, the equation $L \psi^- = 0$
then amounts to
\eqn\ODE{
{d \over dr} \left( \matrix{ \psi^-_s \cr \psi^-_v } \right) = \left( \matrix{
-2 A - \phi & -{\rm Re} \, m \cr -{\rm Re} \, m & -2 r^{-1} + 2 A - \phi}
\right) \left( \matrix{ \psi^-_s \cr \psi^-_v } \right).
}
This system of coupled linear differential equations has two linearly
independent solutions. However, to have a regular solution at the origin, we
must impose the boundary condition $\psi^-_v (0) = 0$. For large $r$, this
solution will generically have components along both eigenvectors of the matrix
in \ODE, so to have a normalizable solution we must require that the
eigenvalues $\lambda_1 = -\sqrt{u} - {\rm Re} \, m$ and $\lambda_2 = -\sqrt{u}
+ {\rm Re} \, m$ in the $r \rightarrow \infty$ limit are both negative. This
will be the case unless
\eqn\condition{
{\rm Re} \, {m \over \sqrt{u}} > 1,
}
where we have stated the condition in a $U(1)_{\cal R}$ invariant way. For
fixed $m$, this inequality determines a heart-shaped region in the finite
$u$-plane where the quark-soliton bound state is absent.

Finally, we briefly comment on the quantum numbers of the soliton states. In
the region of moduli space where we have two such states, their quark number
charges $S$ differ by the quark number charge of the elementary charge, i.e. by
one unit. Furthermore, for $m = 0$, the theory is invariant under quark number
conjugation, which fixes the quark numbers of the soliton states to $S = -1/2$
and $S = 1/2$ \JR. For non-zero $m$, the quark number of the lowest soliton
state has been calculated in \ref\GW{
J. Goldstone and F. Wilczek, {\it Phys. Rev. Lett.} {\bf 47} (1981) 986.
}, and $S \rightarrow 0$ as $m \rightarrow \infty$. At weak coupling, the
decrease of $S$ with increasing $m$ is rapid.

\bigskip I am grateful to the Aspen Center for Physics for its hospitality and
to G. Moore for discussions. This research was supported by DOE under grant
DE-FG02-92ER40704.

\listrefs

\bye